\documentclass{ws-procs9x6}

\begin{document}

\title{Study of the gluon polarization in the proton with a silicon vertex upgrade at RHIC/PHENIX}

\author{M.Togawa for the PHENIX collaboration}

\address{Kyoto University, Kyoto, 606-8502, JAPAN. \& \\ 
  Radiation Laboratory, RIKEN, Wako, Saitama, 351-0198, JAPAN. \\
  E-mail: togawa@nh.scphys.kyoto-u.ac.jp}

\maketitle

\vspace{-5mm}

\abstracts{
  \textbf{Abstract.}
  PHENIX has a well defined program for measuring the polarized gluon distribution in the nucleon.
  The measurements of gluon polarization $via$ the direct-photon production and the heavy-flavor production can be significantly improved by the silicon vertex tracker upgrade.   
  We have studied the possible improvements of the gluon polarization measurements using Monte Carlo simulation and they are shown and discussed in this paper.
}

\vspace{-8mm}

\section{Introduction}

 One of the major goals of the PHENIX experiment at the Relativistic Heavy Ion Collider (RHIC) is a determination of the gluon polarization in the proton \cite{1}.
 We have measured the double longitudinal-spin asymmetry $A_{LL}$ of $\pi^{0}$ production in longitudinally polarized $p$-$p$ collisions at $\sqrt{s}$ = 200 GeV \cite{2}.
 In this process, $\pi^{0}$ carries only fraction of the momentum of the scattering quark or gluon.
 More direct information of the gluon polarization will be obtained by measuring $A_{LL}$ of direct-photon production and heavy-flavor production at $\sqrt{s}$ = 200 GeV and 500 GeV.
 Sensitivity of the measurements will be improved by using a silicon vertex tracker (VTX) which is proposed as a PHENIX upgrade plan \cite{3}.
 It consists of four layers of barrel detectors as shown in Fig.1, and covers 2$\pi$ azimuthal angle and pseudo-rapidity $|\eta|<$1.2.
 Inner two layers are silicon pixel detectors and outer two layers are silicon strip detectors \cite{4}.
 In this paper, we will discuss improvements of the gluon polarization measurements using Monte Carlo simulations.

\begin{figure}[htp]
\begin{center}

  \begin{minipage}{40mm}
  \begin{center}
    \includegraphics[scale=0.25]{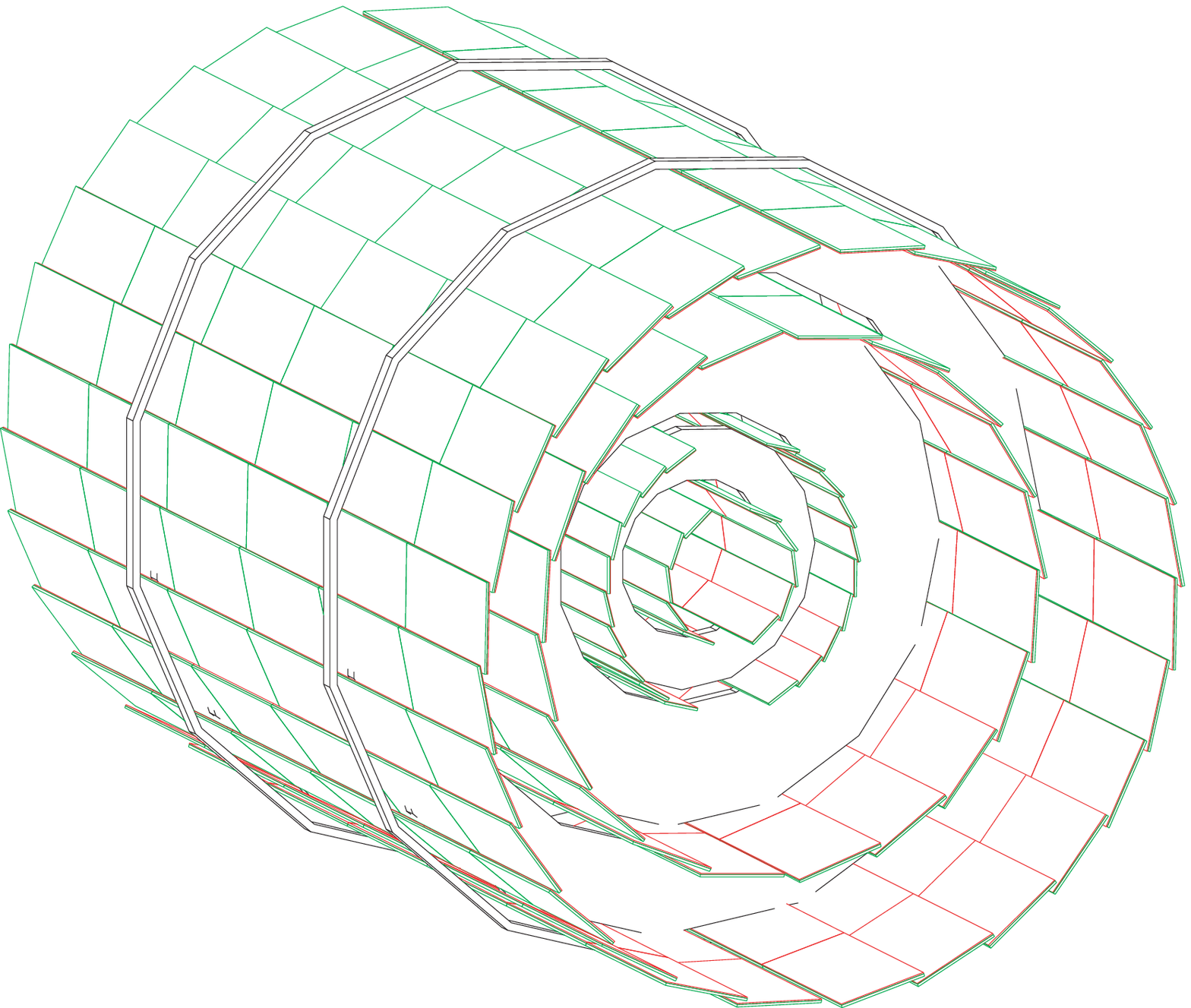} \\
  \end{center}
  \end{minipage}
  \hspace{1mm}
  \begin{minipage}{70mm}
  \begin{center}
      \begin{tabular}{ c || c | c | c } \hline
	
	\  Layer & Si & Radiation & Radius (cm)\\ 
	\  & & length & \\ \hline \hline
	\  1 & pixel & $\sim$ 1 \% & 2.5 \\  
	\  2 & pixel & $\sim$ 1 \% & 5.0 \\  
	\  3 & strip & $\sim$ 2 \% & 10.0 \\  
	\  4 & strip & $\sim$ 2 \% & 14.0 \\ \hline
	
      \end{tabular}

      Pixel size (50$\mu$m$\times$425$\mu$m) \\
      Strip size (80$\mu$m$\times$3cm) \\
  \end{center}
  \end{minipage}


  \label{Barrel}
  \caption{A view of Barrel part of VTX. Parameters are written in table.}

\end{center}
\end{figure}

\section{Direct-photon measurement}\label{subsec:dp}

 Direct-photon process is dominated by the gluon Compton process, whose contribution to the $A_{LL}$ measurement is written as :
 \begin{equation} 
   A_{LL}^{qg \to q\gamma}(p_{\rm{T}}) \sim \frac{\Delta g(x_{g})}{g(x_{g})} \cdot
   A_{1}^{p}(x_{q}) \cdot \hat{a}_{LL}^{qg \to q\gamma }
   \label{eq1}
 \end{equation}
 where $A_{1}^{p}(x_{q})$ shows the quark polarization measured by polarized-DIS experiments, and $\hat a_{LL}$ is the calculable partonic level asymmetry of the gluon Colmpton process.
 $x_{g}$ and $x_{q}$ are momentum fractions of the gluon and quark, respectively.
 The gluon polarization, $\Delta g(x_{g})/g(x_{g})$, can be extracted with this formula.

 Direct-photons are detected by two Central Arms in the current PHENIX setup.
 A central arm is composed of a Electro-Magnetic Calorimeter (EMCal) system, a tracking system and the particle-ID detectors .
 It covers 90-degree azimuthal angle and $|\eta|<$0.35. 
 $x_{g}$ and $x_{q}$ are evaluated to be $\langle x_{g} \rangle$ = $\langle x_{q} \rangle$ = 2 $p_{\rm{T}}/\sqrt{s}$ with photon's $p_{\rm{T}}$ as approximate average values.

 With the VTX, we will be able to also identify $jets$ by charged-particle tracking with expanded acceptance.
 Reconstruction resolution of the $jet$-axis is estimated to be 0.13 in RMS by PYTHIA simulation studies.
 By detecting both the direct-photon and $jet$ in one event, $x_{g}$ and $x_{q}$ can be evaluated under two assumptions, ($a$) $p_{\rm{T}}$ balance of the photon and $jet$, ($b$) $x_{g}$ is smaller than $x_{q}$:

 \begin{eqnarray} 
   x_{1} = \frac{p_{\rm{T}}}{\sqrt{s}}(e^{\eta_{jet}}+e^{\eta_{\gamma}}) \hspace*{20pt}
   x_{2} = \frac{p_{\rm{T}}}{\sqrt{s}}(e^{-\eta_{jet}}+e^{-\eta_{\gamma}}), \\
   x_{g} = \min\{x_{1},x_{2}\} \hspace*{20pt} x_{q} = \max\{x_{1},x_{2}\} \notag
   \hspace{10mm}
   \label{eq2}
 \end{eqnarray}
 where $\eta_{jet}$ and $\eta_{\gamma}$ are pseudo-rapidity of the $jet$-axis and direct-photon, respectively.
 Figure 2 shows comparison of $x_{g}$ given by PYTHIA and the reconstructed $x_{g}$.
 Better correlation can be found in the right figure.
 It indicates determination of the $x_{g}$ improves with the VTX upgrade.

\begin{figure}[htp]
\begin{center}

  \begin{minipage}{50mm}
  \begin{center}
    \includegraphics[scale=0.25]{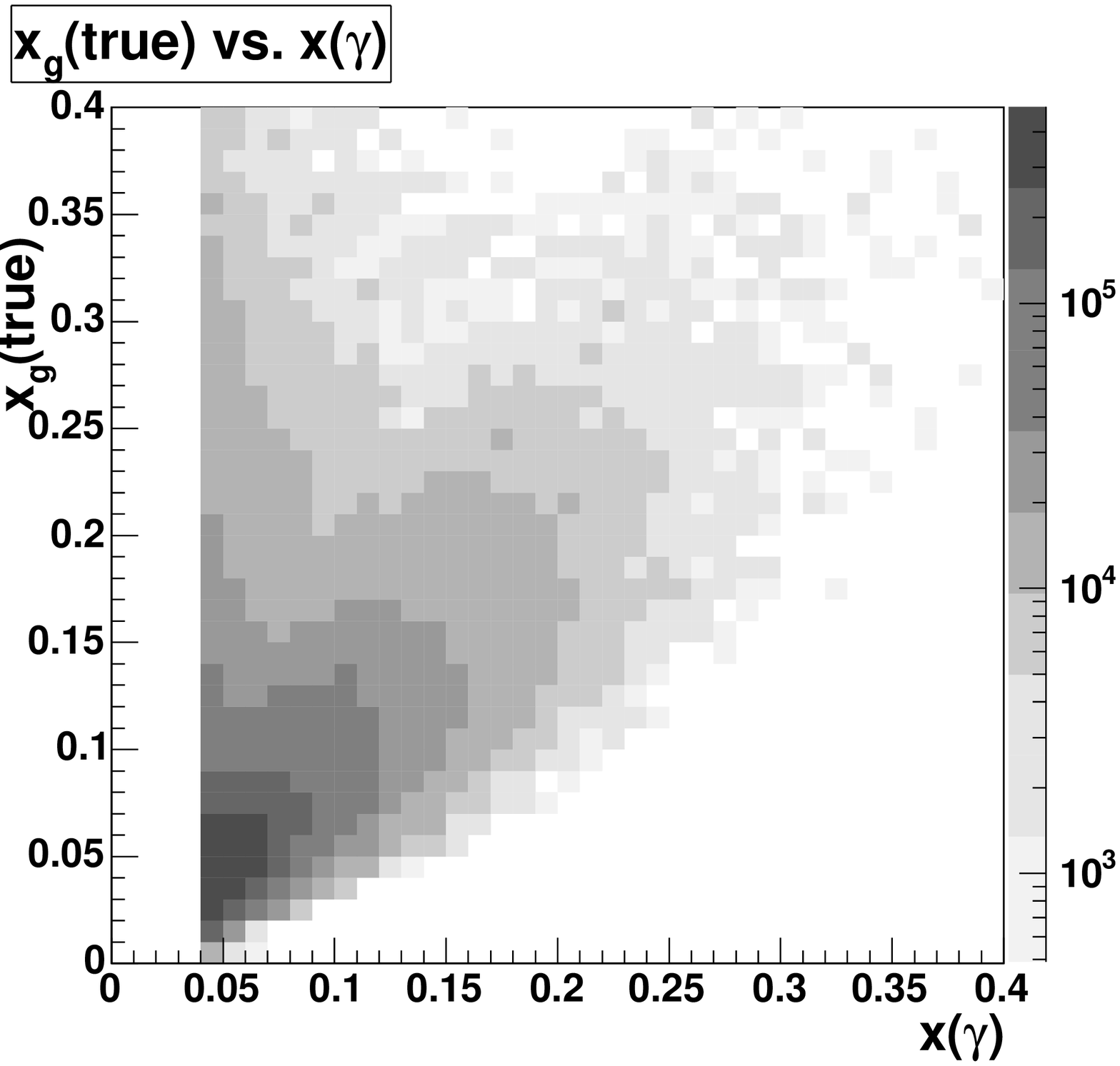}
    (a) : measurement only photon
  \end{center}
  \end{minipage}
  \hspace{1mm}
  \begin{minipage}{50mm}
  \begin{center}
    \includegraphics[scale=0.25]{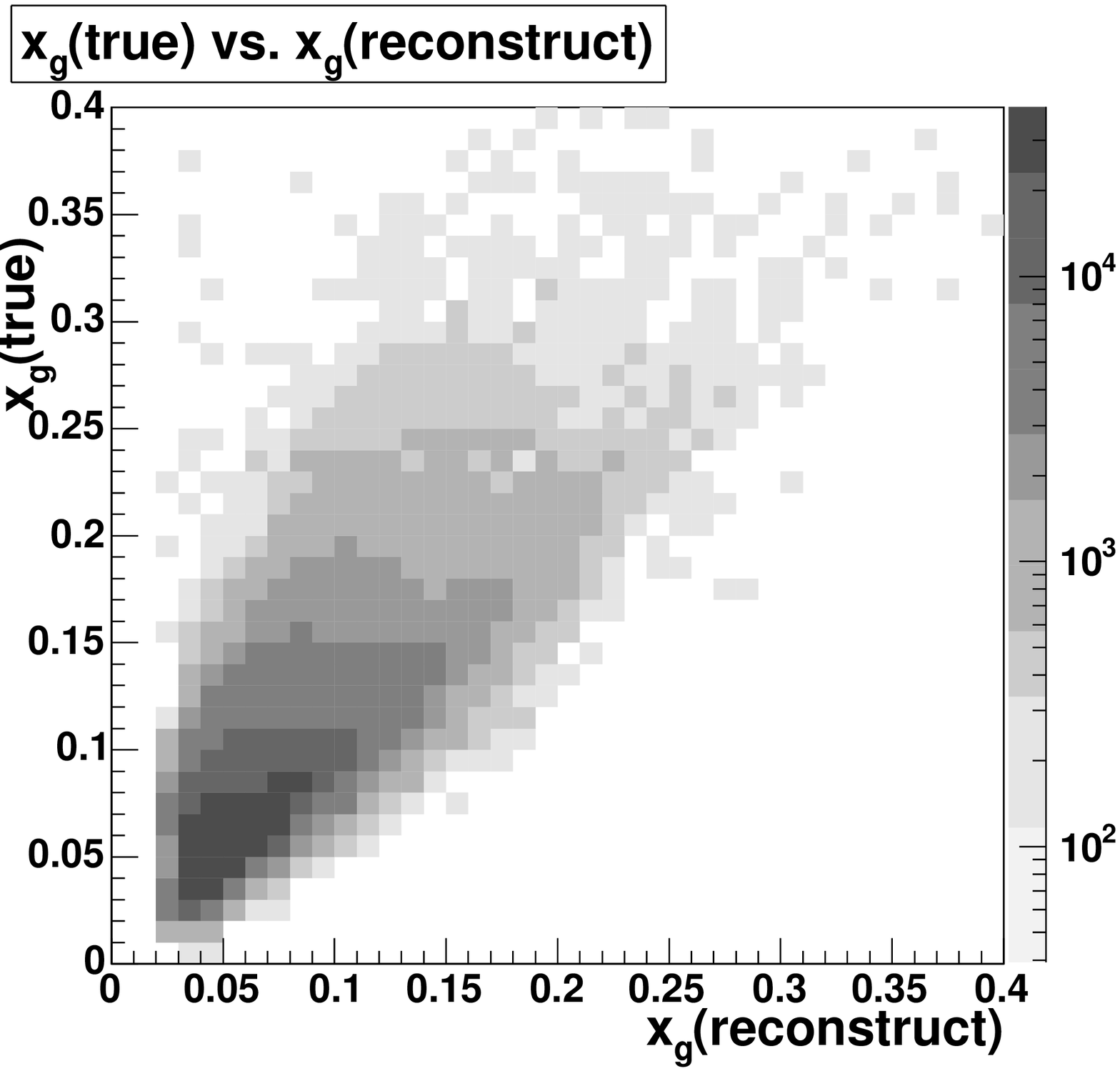}
    (b) : with $jet$-axis using VTX 
  \end{center}
  \end{minipage}

  \caption{Comparizon of $x_{g}$ from PYTHIA and reconstructed $x_{g}$. (a) : $x_{g}$ is calculated using $p_{\rm{T}}$ of photon (current setup). (b) : $x_{g}$ is calculated using equ.(2) with the VTX.}
  \label{CompareX}

\end{center}
\end{figure}

\section{Single-electron measurement with a displaced vertex}

 In PHENIX, we measure $D$- and $B$-meson production $via$ their decays to the single electron.
 Electrons are detected by the Central Arms, and well identified by ring-imaging Cherenkov detectors and EMCal \cite{5}.

 Heavy-flavor production is dominated by the gluon fusion process.
 $A_{LL}$ of the gluon fusion process is shown as: 
 \begin{equation} 
   A_{LL}^{gg \to Q \bar Q} \sim \frac{\Delta g(x_{1})}{g(x_{1})} \cdot
   \frac{\Delta g(x_{2})}{g(x_{2})} \cdot
   \hat{a}_{LL}^{gg \to Q \bar Q}
   \label{eq3}
 \end{equation}
 Since the $\hat a_{LL}^{gg \to Q \bar Q}$ value depends on heavy-quark mass, $A_{LL}^{gg \to c \bar c}$ and  $A_{LL}^{gg \to b \bar b}$ are different \cite{6}.
 To extract the gluon polarization from the single-electron measurement, we need to distinguish these two origins.

 Since $D$- and $B$-meson have long and different life time, they can be distinguished by measuring a Distance of Closest Approach (DCA) of the single electron track from the primary vertex.
 The VTX can determine them with a vertex resolution less than 50 $\mu$m .
 We studied a performance of the VTX to distinguish single electrons from $D$-meson, $B$-meson and Dalitz dacays by applying a DCA cut with PYTHIA which is tuned to reproduce charm and beauty production at FNAL fixed target experiments and CDF \cite{3}.
 Yield of electrons without the DCA cut, $N_{e}^{no}$\hspace{0.5mm}$^{cut}$, and that with the DCA cut, $N_{e}^{cut}$, can be presented : 
 \begin{eqnarray} 
   N_{e}^{no}\hspace{0.5mm}^{cut} &=& N_{e}^{\rm{Dalitz}} + N_{e}^{D} + N_{e}^{B} \\
   N_{e}^{cut} &=& R^{\rm{Dalitz}}\cdot N_{e}^{\rm{Dalitz}} + R^{D}\cdot N_{e}^{D} + R^{B}\cdot N_{e}^{B} \notag
 \end{eqnarray}
 $R$ is an efficiency of each decay for the DCA cut, which can be estimated by the Monte Carlo simulation with the life time.
 Yield of electrons from the Dalitz decay ($N_{e}^{\rm{Dalitz}}$) can be evaluated well with measured $\pi^{0}$ data, thus we can decompose yield of electrons from $D$-meson ($N_{e}^{D}$) and that from $B$-meson ($N_{e}^{B}$).
 Figure 3 shows estimated errors of decomposed $A_{LL}$ of single electrons from $D$-meson decays and that from $B$-meson decays assuming luminosity $L$ = 320 pb$^{-1}$ and polarization of protons are 70\% at $\sqrt{s}$ = 200 GeV.

\begin{figure}[htp]
\begin{center}
    \includegraphics[scale=0.35]{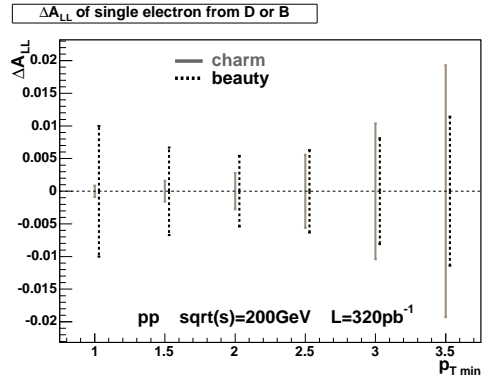}
    \caption{The $A_{LL}$ errors of electrons from charm and beauty. Each point shows $A_{LL}$ error of the $p_{\rm{T}}$ region above $p_{\rm{T}min}$.}
    \label{AllErrorHeavy}
\end{center}
\end{figure}

\vspace{-10mm}

\section{Conclusion}
 We have studied improvement of the gluon polarization measurement in the proton with the VTX upgrade at PHENIX.
 Determination of $x_{g}$ in direct-photon production will be improved by measuring the $jet$-axis with the VTX.
 In heavy-flavor production, single electrons from $D$- and $B$-meson decays will be distinguished by measuring the DCA with precise determination of their track and the primary vertex of the collision.
 The VTX will be installed in summer of 2007, and will be operated in 2007-2008 RUN.

\end{document}